\documentclass[twocolumn]{article}

\usepackage{authblk}
\usepackage{hyperref}
\usepackage[margin=2cm]{geometry}
\usepackage{graphicx}
\usepackage{dcolumn}
\usepackage{bm}
\usepackage{physics}
\usepackage{authblk}

\usepackage[utf8]{inputenc}
\usepackage[T1]{fontenc}
\usepackage{mathptmx}
\usepackage{etoolbox}

\usepackage{graphicx}
\usepackage{dcolumn}
\usepackage{bm}
\usepackage{physics}
\renewcommand{\vec}{\mathbf}
\usepackage{chemformula}

\usetikzlibrary{shapes,arrows,arrows.meta}
\tikzstyle{box} = [rectangle, rounded corners, minimum width=3cm, minimum height=1.25cm,text centered, draw=black]
\tikzstyle{round box} = [rounded rectangle, minimum width=3cm, minimum height=1.5cm,text centered, draw=black]
\tikzstyle{bigbox} = [rectangle, rounded corners, minimum width=4cm, minimum height=2cm,text centered, draw=black]
\tikzstyle{arrow} = [thick,->,>=stealth]
\tikzstyle{doublearrow} = [<->,>=stealth, thick, style=double]

\providecommand{\keywords}[1]
{
  \small	
  \textbf{\textit{Keywords---}} #1
}

\begin{document}

\title{
Density functionals with spin-density accuracy for open shells
}

\author{Timothy Callow$^{1,2,3,4}$, Benjamin Pearce$^1$, Nikitas Gidopoulos$^1$ \\
\small $^1$Department of Physics, Durham University, South Road, Durham, DH1 3LE, UK \\
$^2$ Max-Planck-Institut f\"ur Mikrostrukturphysik, Weinberg 2, D-06120 Halle, Germany \\
$^3$ Center for Advanced Systems Understanding (CASUS), D-02826 G\"orlitz, Germany \\
$^4$ Helmholtz-Zentrum Dresden-Rossendorf, D-01328 Dresden, Germany 
} 

\date{}

\twocolumn[
  \begin{@twocolumnfalse}
    \maketitle
\begin{abstract}
Electrons in zero external magnetic field can be studied with density functional theory (DFT) or with spin-DFT (SDFT). 
The latter is normally used for open shell systems because its approximations appear to model better the exchange and correlation (xc) functional, 
but also because so far 
the application of DFT implied a closed-shell-like approximation. 
Correcting this error for open shells allows the approximate DFT xc functionals to become as accurate as those in SDFT. 
In the limit of zero magnetic field, the Kohn-Sham equations of SDFT emerge as the generalised KS equations of DFT.
\end{abstract}
  \end{@twocolumnfalse}
]


\section{Introduction}

Density functional theory (DFT) \cite{HK64,kohnsham} and spin-density functional theory (SDFT) \cite{Hedin_barth,gidopoulos2007SDFT} 
are two reformulations of the many-electron problem. Their computational advantage is replacing the solution of Schr\"odinger's equation 
for the multi-dimensional ground state of the physical electronic system of interest by the solution of a much less demanding equation 
that yields the ground-state (total) density $\rho$ (DFT) or the spin-density $(\rho^\uparrow, \rho^\downarrow)$ (SDFT) of the real 
system. For simplicity, we discuss the special case of SDFT for a collinear magnetisation density 
$m=-\mu_0 (\rho^\uparrow - \rho^\downarrow)$; the generalisation to the noncollinear case is straightforward.

Both theories can be applied equally well to study electronic systems when the external magnetic field, $B$, vanishes. Then, obviously, the predicted values for any observable quantity must be the same between the two, formally exact, theories. Specifically, the ground-state (total) density and energy of the electronic system must be the same in both theories. 

In DFT/SDFT the value of any observable quantity becomes a functional of the density/spin-density, but so far little is known about the functional dependence of general observable quantities, other than the energy and identically the total density (DFT) or the spin-density (SDFT). For example, the spin-density of the exact SDFT-KS system is equal to the true spin-density of the real system. However, this does not hold for the spin-density of the exact DFT-KS system when $B=0$. In order to obtain the real (observable) spin-density from the density, we need the so far unknown density functional for the spin-density.

In the absence of a magnetic field, $B=0$, by definition, the virtual KS systems in DFT and SDFT of non-interacting electrons share with the real (interacting) system the same total density (DFT and SDFT) and the same spin-density (SDFT). For open shell systems, although the two KS systems have equal total densities, they are not the same and the single-particle KS equations of exact SDFT do not reduce to those of exact DFT. 

On reflection it seems paradoxical that, for $B=0$ the virtual, non-interacting KS system of SDFT does not reduce to the virtual, non-interacting KS system of DFT, and that the two have different spin-densities when both share the same total density. The apparent paradox is related to the limit of the SDFT single-particle equations when $B=0$ and is addressed in the second part of this paper. 

Even though for $B=0$ both exact theories can equally well describe an electronic system, this changes when density functional approximations (DFAs) are introduced. The flexibility of two basic variables in SDFT is thought to offer a better modelling of the exchange-correlation energy, $E_\textrm{\rm xc} [\rho^\uparrow , \rho^\downarrow]$, compared with just a single variable in DFT, $E_\textrm{\rm xc} [\rho] $. As Parr and Yang write \cite{Parr_Yang}, in local and semi-local density functional approximations ``the exchange-correlation energy of the electrons is approximated locally by results for the homogeneous spin-compensated electron gas. Such a procedure is not appropriate for systems with unpaired electrons, like open-shell molecules. A better description for such systems will be obtained through the use of the exchange-correlation energy of the homogeneous spin-polarised electron gas''. As a result, open shell systems are in practice usually treated in SDFT rather than DFT.

For $B=0$, the main error in modelling open shell electronic systems as unpolarised arises in the exchange energy functional, $E_\textrm{x} [\rho]$ (DFT) vs $E_\textrm{x} [\rho^\uparrow , \rho^\downarrow ]$ (SDFT).
When a local or semi-local approximate exchange energy expression is employed, depending explicitly only on the density $\rho$ and its derivatives, it is essentially assumed that the spin-up and spin-down densities are equal to each other and to one half of the total density. For open shell systems, this amounts to mixing partly the spin-up $\rho^\uparrow$ with the spin-down $\rho^\downarrow$ densities in the exchange energy expression:
\begin{equation}
\label{mixrho}
E_\textrm{x} [ \rho ] \simeq E_\textrm{x} [\rho/2 , \rho/2 ] , \ \ 
\rho = \rho^\uparrow + \rho^\downarrow .
\end{equation} 

This mixing of the spin-densities in the exchange energy leads to a spurious error that we call the ghost exchange energy error, in analogy to the `ghost interaction' error of Ref.~\cite{Ghost_interaction}. We define it by the difference of the exchange energies with mixed spin-densities from the reference exchange energy where the spin-densities are separate, as they must be in the KS Slater determinant: 
\begin{equation}
\label{ghostx}
G_{\rm x} = 
E_{\rm x} [ \rho /2 , \rho /2 ] - 
E_{\rm x} [ \rho^{\uparrow} , \rho^{\downarrow} ] .
\end{equation}

In DFT, the ghost exchange error is not caused by the approximate or exact expression for the exchange energy functional but by the mixing of two spin-densities when an open shell system is treated as if it were closed-shell, considering half its electrons as spin-up and half as spin-down. It follows that the ghost exchange error would even be present for the exact exchange energy density functional, were the relation \eqref{mixrho} to be (incorrectly) assumed. 

In the following first part of this paper, we show that open shell systems in zero magnetic field can already be modelled within DFT, without any mixing of spin-up and spin-down KS orbitals and avoiding the ghost-exchange error. In the literature there are alternative approaches 
for open-shell systems based on a spin-restricted formalism, for example we mention the restricted open-shell KS (ROKS) 
method \cite{Filatov1,Filatov2}. 
This has no ghost exchange error, but uses a linear combination of KS determinants to satisfy symmetry constraints in a similar spirit to the ROHF method \cite{ROHF}. Our approach, on the other hand, requires only a single Slater determinant and is thus formally defined in KS-DFT.


\section{The xc energy as an implicit density functional}

To continue, we consider the exact exchange energy in KS theory. It is given by the Fock expression in terms of the spin-orbitals of the KS Slater determinant,
\begin{equation}
     E_\textrm{x} [\rho^{\uparrow} , \rho^{\downarrow} ] =
     -\frac{1}{2}\iint \!\! \dd{\vec{r}} \dd{\vec{x}} \left\{\frac{|\rho^{\uparrow}(\vec{r},\vec{x})|^2}{|\vec{r}-\vec{x}|} + \frac{|\rho^{\downarrow}(\vec{r},\vec{x})|^2}{|\vec{r}-\vec{x}|} \right\}.
\end{equation}
where $\rho^\sigma (\vec{r}, \vec{x})$, $\sigma=\uparrow, \downarrow$, 
is the spin-$\sigma$, one-body reduced density matrix of KS orbitals.

The exact exchange energy functional separates in two disjoint terms,
\begin{equation} \label{separ}
E_\textrm{x}[\rho^{\uparrow} , \rho^{\downarrow} ] = E_\textrm{x}[\rho^{\uparrow} , 0] + E_\textrm{x}[0 , \rho^{\downarrow} ],
\end{equation}
in which the subsets of spin-up and spin-down KS orbitals do not mix. Hence, in order to avoid cross exchange effects between opposite spin-electrons also in approximations, the approximate exchange energy density functionals must also satisfy equality \eqref{separ}.

Since modelling the approximate exchange energy in terms of the total density and its derivatives, as in LDA and in semi-local DFAs, always violates equality \eqref{separ}, we have to model the exchange energy using the KS spin-density \footnote{In DFT the KS spin-density need not coincide with the true spin-density.}. This modelling is still within DFT (not SDFT) since in DFT the KS spin-density (the spin-density of the KS determinant) is an implicit functional of the total density. 

Equality~\eqref{separ} is satisfied by the local spin-density approximation (LSDA), and as far as we know all spin-dependent density functionals (DFs) for the exchange energy, approximate or the exact one. Therefore, we shall use expressions for the exchange energy which depend on the total density indirectly, or implicitly, via the KS spin-density (spin-density functionals, SDFs), 
\begin{equation}
\label{iDFx}
    E_\textrm{x}^\textrm{iDF}[\rho]=E_\textrm{x}^\textrm{SDF} \big[\rho^{\uparrow}[\rho],\rho^{\downarrow} [\rho] \big],
\end{equation}
where $E_{\rm x}^{\rm SDF}$ must be the sum of two disjoint terms \eqref{separ}. 

The acronyms iDF and SDF denote that the exchange energy on the lhs is an \emph{implicit} density-functional and that the exchange energy expression on the rhs depends on the KS spin-density.

Given that the exchange-functional is now an implicit functional of the density, and that the correlation energy is modelled better in terms of the spin-density, we choose to write also the correlation energy functional as an implicit functional, depending on the total density via the KS spin-density, like exchange \footnote{The correlation energy is not a sum of two disjoint terms and does not satisfy \eqref{separ}}. Indeed in approximations there is a small gain in accuracy in treating correlation using spin-polarised DFs, though we have observed this is typically a minor contribution compared to the ghost exchange error. The whole exchange-correlation (xc) density functional is thus written as
\begin{equation}
\label{iDFxc}
    E_\textrm{\rm xc}^\textrm{iDF}[\rho]=E_\textrm{\rm xc}^\textrm{SDF} \big[\rho^{\uparrow}[\rho],\rho^{\downarrow} [\rho] \big].
\end{equation}

Hence, the total energy DF is now given by (for the exact or approximate DFs)
\begin{equation}
\label{iDFE}
     E_{v_{\rm en}}^\textrm{iDF}[\rho] = T_\textrm{s}[\rho] + \int \dd{\vec{r}} v_\textrm{en}(\vec{r}) \rho(\vec{r}) 
     + U[\rho] + E_\textrm{\rm xc}^\textrm{iDF}[\rho],
\end{equation}
where $U[\rho]$ is the Hartree energy and $v_{\rm en} (\vec{r})$ is the external (electron-nuclear) potential of the interacting system. 

As the xc energy functional is an implicit DF, the xc-potential, given by the functional derivative
\begin{equation}
   v_\textrm{\rm xc} [ \rho ] (\vec{r}) = \fdv{ E_\textrm{\rm xc}^\textrm{iDF}[\rho]  }{\rho(\vec{r})},
\end{equation}
must be determined using the optimised effective potential (OEP) method \cite{Sharp_Horton_OEP,Talman_Shawdwick_OEP},
\begin{multline} \label{eq:oepeqn}
    \int{\dd{\vec{r}'}} \sum_{\sigma} \chi^{\sigma}(\vec{r},\vec{r}') v_\textrm{\rm xc} [\rho] (\vec{r}') = \\
    \int{\dd{\vec{r}'}} \sum_{\sigma} \chi^{\sigma}(\vec{r},\vec{r}') 
  v_{\rm xc}^\sigma \big[ \rho^{\uparrow} [ \rho ]  , \rho^{\downarrow} [ \rho ]  \big] (\vec{r} ' )  ,
  \end{multline}
with
\begin{gather}
v_{\rm xc}^\sigma \big[ \rho^{\uparrow} [ \rho ]  , \rho^{\downarrow} [ \rho ]  \big] (\vec{r} ' ) 
=    \frac{\partial E_{\rm xc}^{\rm SDF} [\rho^{\uparrow} , \rho^{\downarrow} ] }{ \partial \rho^{\sigma} ({\bf r} ')} \Bigg|_{ \rho^{\uparrow} = \rho^{\uparrow} [ \rho ]     \atop \rho^{\downarrow} = \rho^{\downarrow} [ \rho ] }  ,  
\label{partialfd} \\
    \chi^{\sigma}(\vec{r},\vec{r}')  = -2\sum_{i=1}^{N^{\sigma}} \sum_{a=N^{\sigma}+1}^\infty \frac{\phi_i(\vec{r})\phi_a(\vec{r})\phi_i(\vec{r}') \phi_a(\vec{r}')} {\epsilon_a - \epsilon_i} .
\end{gather}

This equation is our first important result. The KS-DFT xc potential $v_{\rm xc}$ is given as the weighted sum of the spin-dependent xc potentials $v_{\rm xc}^\sigma$, with weighting factors the spin-dependent response functions $\chi^\sigma$. We argue that this is the only consistent way (that avoids the ghost-exchange error) to evaluate the KS-DFT xc potential for open shell systems. For closed shell systems, the solution reduces to the familiar KS xc functional derivative. For fully spin-polarised systems (eg spin up, $\rho^\uparrow=\rho, \rho^\downarrow=0$), the spin-down response function vanishes, and the solution of \eqref{eq:oepeqn} reduces to the spin-up xc potential, $v_{\rm xc} [\rho] = v_{\rm xc}^\uparrow [\rho, 0]$.  

\section{Results}

We have implemented these equations in the Gaussian basis set code HIPPO \footnote{For information, contact NL at lathiot@eie.gr.}. For more detail on the computational implementation of the OEP equation please see Refs.~\cite{callow_thesis} and \cite{callow_2020_improving}. Unless otherwise stated, all results use cc-pVTZ orbital bases and uncontracted cc-pVDZ auxiliary bases \cite{Dunning_1,Dunning_2}.

Ground-state energies calculated with our iDF KS method compare favourably with those from SDFT-KS. This is demonstrated in Table \ref{tab:gs_energies} for some atoms and molecules at their equilibrium geometries, using the L(S)DA functional. We emphasise again that we seek only the ground-state energy in our approach, and thus where a state is referred to as `doublet' this simply means
\begin{eqnarray}
     \Delta N = N^{\uparrow} - N^{\downarrow} = 1,
\end{eqnarray}
and likewise for triplet and so on. We also note the effect of the ghost exchange (and correlation) error, which is particularly pronounced for triplet states; we shall now explore some examples in which this error significantly affects results.

\begin{table}[t]
    \centering
    \begin{tabular}{cccc}
           & $E_\textrm{LDA}\ (\textrm{Har})$ & $E_\textrm{iLDA} \ (\textrm{Har})$ & $E_\textrm{LSDA}\ (\textrm{Har})$\\
           \hline
        $\ch{Li}$	                &-7.388721	&-7.398145	&-7.398155 \\
        $\ch{B}$                    &-24.43315  &-24.44669  &-24.44747 \\
        $\ch{N}$                    &-54.12891  &-54.14996  &-54.15110 \\
        $\ch{Na}$	                &-161.6491	&-161.6571	&-161.6572 \\
        $\ch{Si}$\footnotemark[1]   &-288.4640  &-288.4905	&-288.4910 \\
        $\ch{LiH^+}$	            &-7.652062	&-7.685603	&-7.685608 \\
        $\ch{O2}$\footnotemark[1]   &-149.6038  &-149.6383	&-149.6403 \\
        $\ch{OH}$                   &-75.34813	&-75.37077	&-75.37208 \\
        $\ch{NH4}$	                &-56.79800	&-56.80389	&-56.80404 \\
        \hline
        Avg \% diff                 & 0.0827 & 0.00101 & -
    \end{tabular}
    \footnotetext[1]{Triplet state} 
    \caption{L(S)DA ground-state total energies calculated with: 
    (i) standard LDA; (ii) implicit LDA (iLDA); and (iii) spin-LDA (LSDA). All states are doublets unless specified otherwise; auxiliary bases are uncontracted cc-pVTZ.} 
    \label{tab:gs_energies}
\end{table}

\begin{figure}[t]
    \centering
    \includegraphics{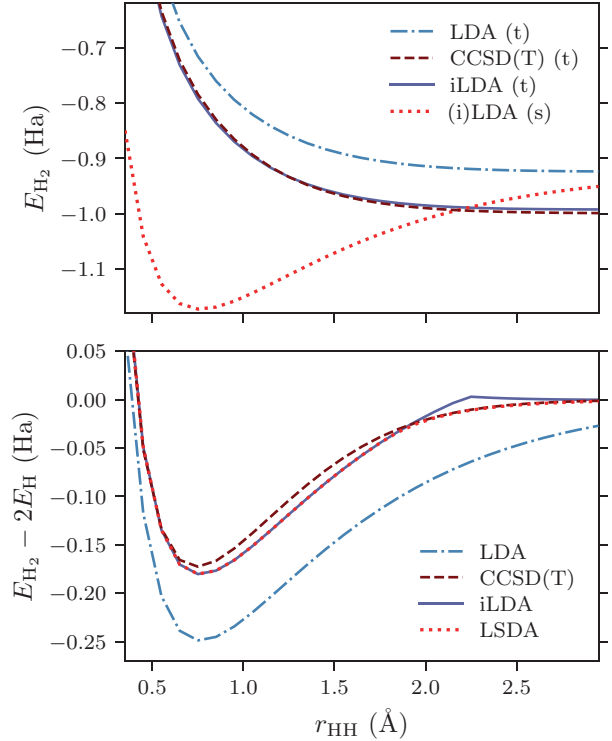}
    \caption{Energy dissociation curves for the \ch{H2} molecule. Top, iLDA energies for different values of $\Delta N$; bottom, comparison of LDA, spin LDA (LSDA), and iLDA minimum energies.}
    \label{fig:H2_dissoc}
\end{figure}

\begin{figure}[t]
    \centering
    \includegraphics{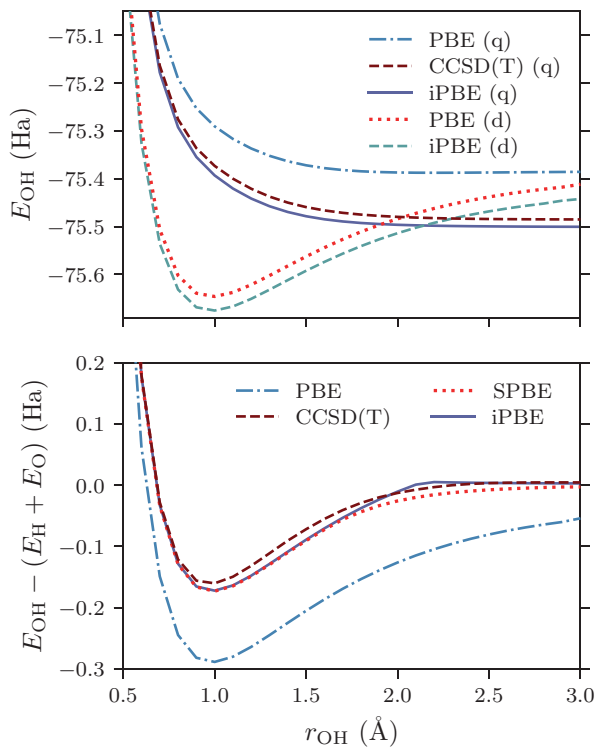}
    \caption{Energy dissociation curves for the \ch{OH} radical. Top, implicit PBE (iPBE) energies for different values of $\Delta N$; bottom, comparison of PBE, spin PBE (SPBE), and iPBE minimum energies.}
    \label{fig:OH_dissoc}
\end{figure}

In Fig~\ref{fig:H2_dissoc}, we have calculated the energy dissociation curve for the \ch{H2} molecule with the L(S)DA functional. As is well-known in the literature \cite{Yang_insights} and can be seen in this figure, the standard restricted solution yields a qualitatively incorrect dissociation curve; however, in our new method, once the bond is stretched enough the triplet state becomes lower in energy than the singlet, and the energy tends to the correct limit. This transition does not occur when the ghost exchange energy is present: in this case the triplet energy is higher than the singlet at all bond distances.

We see a similar picture emerge for the stretched OH radical with the Perdew-Burke-Ernzehof (PBE) \cite{PBE} functional in Fig.~\ref{fig:OH_dissoc}. In this case, the energy dissociation curve is again qualitatively inaccurate for the solution contaminated with ghost-exchange errors; in our method which removes the ghost-exchange error, the energy correctly becomes equal to the energy of the two atoms in the infinitely stretched limit. In both these examples, the transition region is interesting: the unrestricted solution yields a smooth dissociation curve, whereas in our single-determinant method the transition is abrupt. Of course, our method does not correct the ubiquitous inability of typical semi-local functionals to capture static correlation effects \cite{Yang_static_Ec}. 

\section{Limit of SDFT-KS equations \\for $B=0$.}

We now turn our attention to the limit of SDFT when $B=0$ and compare with DFT. We start with the universal internal energy density functional,
\begin{equation} 
\label{Fdf}
F [ \rho ] = \min_{\Psi \rightarrow \rho} \langle \Psi | {\hat T} + {\hat V}_{\rm ee} | \Psi \rangle   \, .
\end{equation}

The minimising state $\Psi_\rho$ depends on the total density $\rho$. When $\rho$ is the ground state density of an interacting electronic system bound by a local potential $v$ (i.e. when $\rho$ is interacting $v$-representable), $\Psi_\rho $ is the corresponding ground state.  

To proceed, we follow Levy \cite{Levy6062} and 
separate the minimisation in \eqref{Fdf} into two separate minimisations with the same minimum,
\begin{equation}
\label{Fmin}
F [ \rho ] = \min_{( \rho^\uparrow, \rho^\downarrow) \rightarrow \rho}  
\bigg[ \min_{\Psi \rightarrow (\rho^\uparrow , \rho^\downarrow ) } \langle \Psi | {\hat T} + 
{\hat V}_{\rm ee} | \Psi \rangle \bigg] \,.
\end{equation}
The inner minimisation, performed first, is over all states $\Psi$ with spin-density $(\rho^\uparrow, \rho^\downarrow)$. The outer minimisation is over all spin-densities that add up to the total density $\rho$. The first minimisation inside the brackets on the rhs defines SDFT's internal energy functional; we conclude that the DFT internal energy functional can be obtained from the SDFT functional with an extra optimisation over spin-densities with the same $\rho$
(see also Appendix B in Ref.~\cite{Spin_symmetry_dilemma}),
\begin{equation}
\label{Fsdf}
F [ \rho ] = \min_{( \rho^\uparrow, \rho^\downarrow) \rightarrow \rho} F [\rho^\uparrow , \rho^\downarrow ] \,.
\end{equation}

The two minimisations in \eqref{Fmin} have the same overall minimum as the minimisation in \eqref{Fdf}. Therefore, the minimising spin-density, $(\rho^\uparrow_\rho , \rho^\downarrow_\rho )$, is equal to the spin-density of $\Psi_\rho $ in \eqref{Fdf} and hence equal to the spin-density of the real interacting system, even though the minimisation \eqref{Fsdf} is at fixed total density (not fixed spin-density), i.e. within DFT, not SDFT.

The minimisation in \eqref{Fsdf} is worth investigating further. We invoke SDFT's KS system with spin-density $(\rho^\uparrow, \rho^\downarrow)$, to expand the internal energy functional in the usual way, $F[\rho^\uparrow , \rho^\downarrow ] =T_{\rm s}[\rho^\uparrow , \rho^\downarrow ] +E_{\rm xc}[\rho^\uparrow , \rho^\downarrow ] + U[\rho^\uparrow + \rho^\downarrow ] $. Dropping the superscript SDF from the functionals of SDFT, we obtain
\begin{equation}
\label{gks_df}
F [ \rho ] = \min_{( \rho^\uparrow, \rho^\downarrow) \rightarrow \rho} \Big\{ T_{\rm s} [\rho^\uparrow , \rho^\downarrow ] 
+ E_{\rm xc} [\rho^\uparrow , \rho^\downarrow ] \Big\}
+ U [ \rho ] \,.
\end{equation}

Since every SDFT-KS system has a different spin-density, the minimisation effectively searches over all SDFT-KS systems with common total density $\rho$ and returns that with the correct spin-density $(\rho^\uparrow_\rho, \rho^\downarrow_\rho)$. The KS Slater determinant state is $\Phi_{\rho^\uparrow_\rho, \rho^\downarrow_\rho}$.
The minimising SDFT-KS system depends only on the total density $\rho$. The corresponding SDFT-KS potential (functional of the spin-density) 
at the specific spin-density $(\rho^\uparrow_\rho, \rho^\downarrow_\rho)$ is also an implicit functional of $\rho$. 
In the Supplementary Material we show how this unrestricted KS potential emerges directly from the minimisation in \eqref{gks_df}.

The SDFT-KS system that minimises \eqref{gks_df} defines a new non-interacting system in DFT (when $B=0$) represented by the state $\Phi_{\rho^\uparrow_\rho, \rho^\downarrow_\rho}$. We call it the generalised KS (GKS) system, as its derivation is analogous to the well-known GKS scheme in the literature \cite{GKS}.

We conclude that the SDFT-KS equations reduce to the DFT-GKS equations for $B=0$. The DFT-GKS Slater determinant state $\Phi_{\rho^\uparrow_\rho, \rho^\downarrow_\rho}$ gives not only the true total density but also the true spin-density (or magnetisation density). The elusive exact density functional for the spin-density is the spin-density of the DFT-GKS system. 

\section{The spin-symmetry dilemma}

One of the oft-cited issues with SDFT in the absence of an external magnetic field is the spin-symmetry dilemma, so-called because SDFT approximate results yield either accurate total energies but with a poor (broken symmetry) prediction for the spin-density, or an accurate prediction for the spin-density with poor total energies \cite{Spin_symmetry_dilemma,Gunnarson_Lundqvist,Gorling_symmetry}. 
The dilemma lies in that the spin-density is the key quantity and hence SDFT (exact) results are supposed to yield both the exact total energy and the exact spin-density. For example, $\ch{H}_2$ should dissociate into two Hydrogen 
atoms with zero net magnetisation (difference of spin densities): indeed the lowest energy solution of local and semi-local approximations in SDFT gets the energy right, but it wrongly dissociates the molecule into two atoms of opposite 
spin (broken symmetry solution). There is also a self-consistent solution where the spin-densities are correctly equal to each other but the total energy is too high.

Perdew, Savin and Burke \cite{Spin_symmetry_dilemma} do not consider that the spin-symmetry dilemma results from the approximation and to address the issue 
they reinterpret exact SDFT for $B=0$.
%
They write that the KS spin-densities ``are not physical spin densities, but are only intermediate objects (like the Kohn-Sham orbitals...''. 
Instead, they argue that the on-top electron pair density $P(\vec{r},\vec{r})$, alongside the total density, are the two fundamental variables of the theory. 
Since the KS spin-density is no longer a basic variable of 
SDFT \cite{Spin_symmetry_dilemma}, its value is not a 
prediction for the real spin-density of the interacting 
system. 
We comment that considering the SDFT-KS spin-density to give the observable spin-density for any value of 
$B$ except at $B=0$, where it needs re-interpretation according to Ref.~\cite{Spin_symmetry_dilemma}, 
suggests that SDFT may have a singular limit at $B=0$.

We share the principle of Ref.~\cite{Spin_symmetry_dilemma}, namely that the spin-density is no longer the basic quantity in the absence of a magnetic field; 
but in our approach, there is no need to consider the auxiliary on-top pair density, with just the total density being the fundamental variable: 
our (ghost-exchange error
free) theory yields the correct ground-state energy and the correct total density in the dissociation limit, and thus the spin-symmetry dilemma is weakened as a
fundamental problem, since the KS spin-density is not expected equal to the observable spin-density. 

However, we also propose that the DFT-GKS spin-density is the exact density functional of the observable spin-density; knowledge of the (exact) SDFT xc functional $E_{\rm xc}[\rho^\uparrow, \rho^\downarrow]$ is necessary to obtain from the  density the (exact) DFT-GKS spin-density. 
The value of the latter is the same as the SDFT-KS spin-density \footnote{Hence SDFT does not have a singular limit at $B=0$.} 
which, in local and semi-local approximations, demonstrates the spin-symmetry dilemma. 
We believe the explanation is not a singular limit of SDFT when $B=0$ but that the approximate DFT-GKS spin-density is not a sufficiently accurate density functional for the real spin-density. 
According to Cohen, Sanchez and Yang \cite{,Yang_static_Ec,Yang_insights} on the fractional spin error, 
the crux of the problem is the degeneracy of different Slater determinants with the same energy and total density but different spin-densities. 
This kind of strong correlation cannot be captured with local or semi-local functionals.

\section{Conclusion}

We have addressed two related conceptual or fundamental theoretical questions in DFT and spin-DFT.
The first is the challenge how to apply the KS equations of DFT for open shells, avoiding a serious  qualitative error, which we called the "ghost-exchange error". 
Associated with this problem is the widespread belief that spin-DFT approximations are  inherently more flexible and hence can model more accurately the xc energy for open shells than the 
corresponding approximations  of DFT. In the literature this common view seemed plausible, almost obvious. 
We demonstrate that actually density functionals can be  as accurate as spin-density functionals, provided the ghost-exchange error  is corrected.

For zero external magnetic field (B=0), both DFT and spin-DFT can obviously be applied  to study open shell electronic systems and both exact theories should give  identical results for any observable quantity. 
Then, intuitively, one would  expect in the limit $B=0$ the single particle  KS equations of spin-DFT to  reduce to a set of single-particle equations of DFT. 
We demonstrate this intuition is correct and that the SDFT equations reduce to the generalised KS equations of DFT.

\section*{Acknowledgments}
We thank N. Lathiotakis for help with the code HIPPO. We acknowledge enlightening discussions with K. Burke, E.K.U. Gross and N. Lathiotakis.
N.I.G. acknowledges financial support from The Leverhulme Trust, through a Research Project Grant with number RPG-2016-005. 
B.P. acknowledges financial support from EPSRC with grant number EP/R513039/1.\\
During the final stage of the project T.C. was funded by the 
Center for Advanced Systems Understanding (CASUS) which is financed by Germany’s Federal Ministry of Education and Research (BMBF) and by the Saxon Ministry for Science, 
Culture and Tourism (SMWK) with tax funds on the basis of the budget approved by the Saxon State Parliament.

\section*{Data Availability}

The data that support the findings of this study are available from the corresponding author upon reasonable request.



\end{document}